\documentclass[prl,superscriptaddress,showpacs,twocolumn,amsfonts,amsmath]{revtex4}
\usepackage[final]{graphicx}

\newlength\figurewidth
\AtBeginDocument{\setlength\figurewidth{.9\linewidth}}

\usepackage{color}
\let\omp\marginpar\relax\def\marginpar#1{\omp{\color{red}#1}}

\begin{document}
\title{Slow Dynamics in Ion-Conducting Sodium Silicate Melts:
  Simulation and Mode-Coupling Theory}
\date{\today}
\def\edin{%
  \affiliation{University of Edinburgh, School of Physics,
   JCMB The Kings Buildings, Mayfield Road, Edinburgh EH9 3JZ, Scotland}}
\def\mainz{%
  \affiliation{Institut f\"ur Physik, Johannes-Gutenberg-Universit\"at Mainz, 
   Staudinger Weg 7, D-55099 Mainz, Germany}}
\author{Th.~Voigtmann}\edin
\author{J.~Horbach}\mainz

\begin{abstract}
A combination of molecular-dynamics (MD) computer simulation and
mode-coupling theory (MCT) is used to elucidate
the structure--dynamics relation in sodium-silicate melts (NS$x$) of
varying sodium concentration. Using only the partial static structure
factors from the MD as an input, MCT reproduces the large separation in
relaxation time scales of the sodium and the silicon/oxygen components.
This confirms the idea of sodium diffusion channels which are reflected by
a prepeak in the static structure factors around 0.95\,\AA$^{-1}$,
and shows that it is possible to explain the fast sodium-ion
dynamics peculiar to these mixtures using a microscopic theory.
\end{abstract}

\maketitle

Glass-forming mixtures of $\text{Si}\text{O}_2$ with an alkali
oxide such as $\text{Li}_2\text{O}$, $\text{Na}_2\text{O}$,
or $\text{K}_2\text{O}$ are of general interest for the study of
transport mechanisms in amorphous condensed matter. They all show
alkali ion mobility that is much higher, often by orders of magnitude,
than that of the silicon and oxygen atoms forming a tetrahedral network
structure. More than 20 years ago, the idea of preferential diffusion
pathways in the Si-O network was proposed as an explanation for the
fast transport of alkali ions \cite{angell82,greaves85}.  A recent
study of sodium silicate melts using a combination of molecular
dynamics (MD) computer simulations and inelastic neutron scattering
\cite{meyer04} has shown that indeed percolating sodium-rich channels
are formed in the intermediate-range static structure.  The sodium
ions perform a hopping motion along those channels \cite{horbach02},
and simulation studies on different alkali silicates have identified
various properties of the characteristic alkali ion sites involved in
this motion \cite{oviedo98,jund01,cormack02,lammert03}.

If the presence of intermediate-range order in the static structure is
intimately related to the high mobility of the alkali ions, it should be
possible to predict the dynamics, in particular the time-scale separation
between the sodium transport and the silicon/oxygen ``matrix'' dynamics,
from this static structure. Such a structure--dynamics relation is,
however, not easily formulated in terms of a microscopic theory.  Here,
we discuss a theoretical framework in which this is possible.

The mode-coupling theory of the glass transition (MCT)
\cite{goetze91}, has been formulated to provide a
first-principles description of slow dynamics in glass-forming
liquids. It needs as input the static equilibrium information about
a system, in its most common form only the partial static structure
factors $S_{\alpha\beta}(q)$.  The MCT equations then yield dynamic
correlation functions, in particular the normalized dynamic structure
factors, $F_{\alpha\beta}(q,t)=S_{\alpha\beta}(q,t)/S_{\alpha\beta}(q)$
(also called the coherent intermediate scattering functions), and the
incoherent self-intermediate scattering functions of a given species
$\alpha$, $F_\alpha^s(q,t)$.  They can be directly compared to scattering
experiments or simulation results and determine, via Green-Kubo relations,
the transport coefficients.

The programme of predicting, using MCT, dynamical features from the
equilibrium structure of the liquid and testing the predictions against MD
simulations, has proven successful for some paradigmatic dense liquids like
the hard-sphere system \cite{voigtmann04}, binary mixtures of hard spheres
\cite{foffi04}, Lennard-Jones particles \cite{nauroth97,kob02}, a model
of $\text{Ni}\text{Zr}_4$ \cite{mutiara01}, and models of more complicated
dense glass formers \cite{fabbian99b,theis00,chong04}. In these systems, the
agreement was qualitative and often even quantitative. For a computer-modeled
silica melt, MCT was able to predict the form factor of the glass, i.e.\ the
height of the plateau in the dynamic structure factors, qualitatively, and
even to within a few percent if in addition to $S_{\alpha\beta}(q)$ static
triplet correlations were taken into account \cite{sciortino01}.

In this work, the MCT equations are solved for sodium silicates,
using MD-simulated static structure factors as an input.
The resulting dynamic structure
factors are compared to the ones obtained from MD.  We demonstrate
that, at least on a qualitative level, the theoretical framework of MCT
allows to describe the large time-scale separation in sodium silicates.
This establishes a structure-dynamics relation based on static pair
correlations. It confirms that the high mobility of sodium ions in sodium
silicates is indeed intimately related the presence of intermediate
range order (i.e.\ the aforementioned channel structure) as reflected by
a prepeak in partial static structure factors.

MD simulations were performed for sodium silicates with different
compositions, $(\text{Na}_2\text{O})(\text{Si}\text{O}_2)_x$ [NS$x$], with
$x=2$, $3$, $5$, $20$, and $40$. The simulations were all done at constant
density $\rho=2.37\,\text{g}/\text{cm}^3$, using $N=8064$ particles and
a slight modification of the pair potential by Kramer \textit{et~al.}\
\cite{kramer91}, which is based on \textit{ab initio} calculations. For
details, we refer to Ref.~\cite{horbach99c,horbach01b}. Based on
the thus obtained $S_{\alpha\beta}(q)$ for NS2 and NS20, the MCT
equations of motion for $S_{\alpha\beta}(q,t)$ \cite{goetze87b}
were solved numerically as outlined in Ref.~\cite{goetze03}, with
discrete wave numbers $0.1\,\text{\AA}^{-1}\le q<20\,\text{\AA}^{-1}$
of step size $\Delta q=0.1\,\text{\AA}^{-1}$, applying a five-point
running average to the input. The MCT solutions are fully determined
only through $S_{\alpha\beta}(q)$ and the static triplet correlation
function $c^{(3)}_{\alpha\beta\gamma}(\vec q,\vec k)$. We set the latter
to zero thus approximating the three-point correlations by a convolution
approximation.

\begin{figure}
\includegraphics[width=\figurewidth]{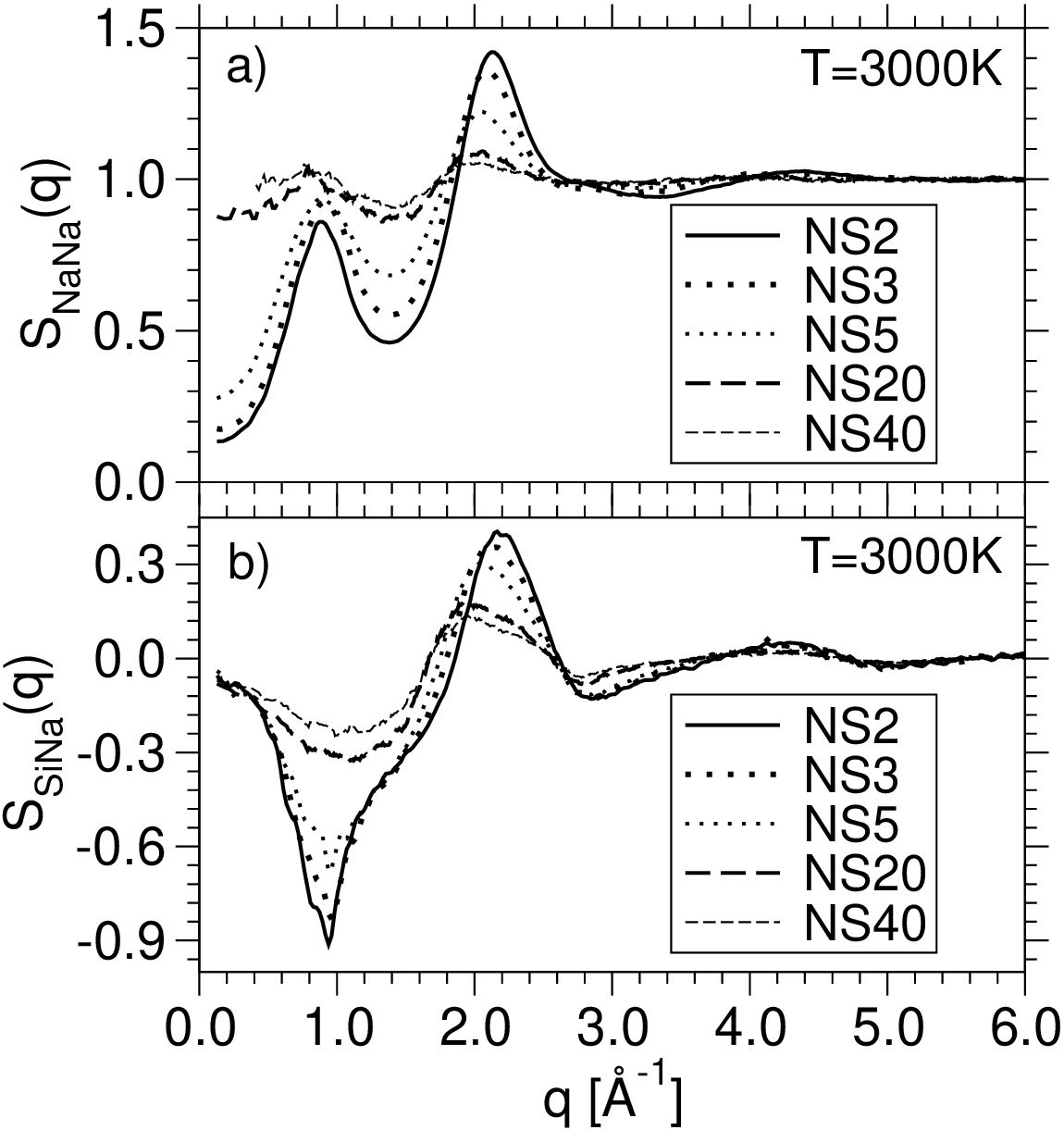}
\caption{\label{fig1}
  Partial structure factors for the indicated sodium silicates at
  $T=3000\,\text{K}$, obtained from MD simulations.
  a) $S_{\text{Na}\text{Na}}(q)$, b) $S_{\text{Si}\text{Na}}(q)$.
}
\end{figure}
The structure factor input is exemplified in Fig.~\ref{fig1}, where we
plot $S_{\text{Na}\text{Na}}(q)$ and $S_{\text{Si}\text{Na}}(q)$ for
different sodium concentrations at the temperature $T=3000\,\text{K}$.
Note that even at this high temperature, typical ``matrix'' relaxation times
are already in the range of 10--100\,ps. 
The channel structure formed by the
alkali ions on intermediate length scales, at about $6$--$7\,\text{\AA}$,
is clearly visible as a prepeak around $q=0.95\,\text{\AA}^{-1}$. This
prepeak becomes more intense with increasing sodium concentration,
and, remarkably, its location does not change significantly over the
whole considered sodium concentration range. This is in agreement
with our recent finding that the prepeak reflects a characteristic
intermediate-range chemical ordering of sodium that leads to the
formation of sodium-diffusion channels.  However, in Fig.~\ref{fig1},
no obvious scale separation hints towards the peculiar sodium dynamics.
In MCT for single-component systems, relaxation times are typically
correlated with peaks in $S(q)$; thus, from a peak in the diagonal
element $S_{\alpha\alpha}(q)$, one might expect a slowing down of the
dynamics around the prepeak position, but nothing particular for the
self-relaxation times.

In the comparison of dynamical quantities between MD and MCT, one has
to note that the theory predicts a divergence of relaxation times at
$T=T_c$: the idealized glass transition, which is smeared out in experiment.
A scaling analysis of the MD data gives a value for $T_c$, which is lower
than the value calculated within the theory.
Hence, in order to compare dynamical data that show the same
degree of slowing down, i.e.\ the same separation between the structural
relaxation time scale and the time scale of molecular short-time motion,
we compare MD data for temperature $T$ with MCT data at a different
(higher) temperature $T_{\text{MCT}}$. This is then the only free
parameter in our comparison.

Note that, despite the quantitative error in the value of $T_c$,
the qualitative change of $T_c$ with varying sodium concentration
is captured by MCT: from a scaling analysis of the simulation data,
one gets $T_c\approx3330\,\text{K}$ for silica \cite{horbach99} and
$2000\,\text{K}$ for NS2 \cite{horbach02b}. The respective values
we calculate within the theory are $T_c\approx 3983\,\text{K}$ and
$3105\,\text{K}$.

\begin{figure}
\includegraphics[width=\figurewidth]{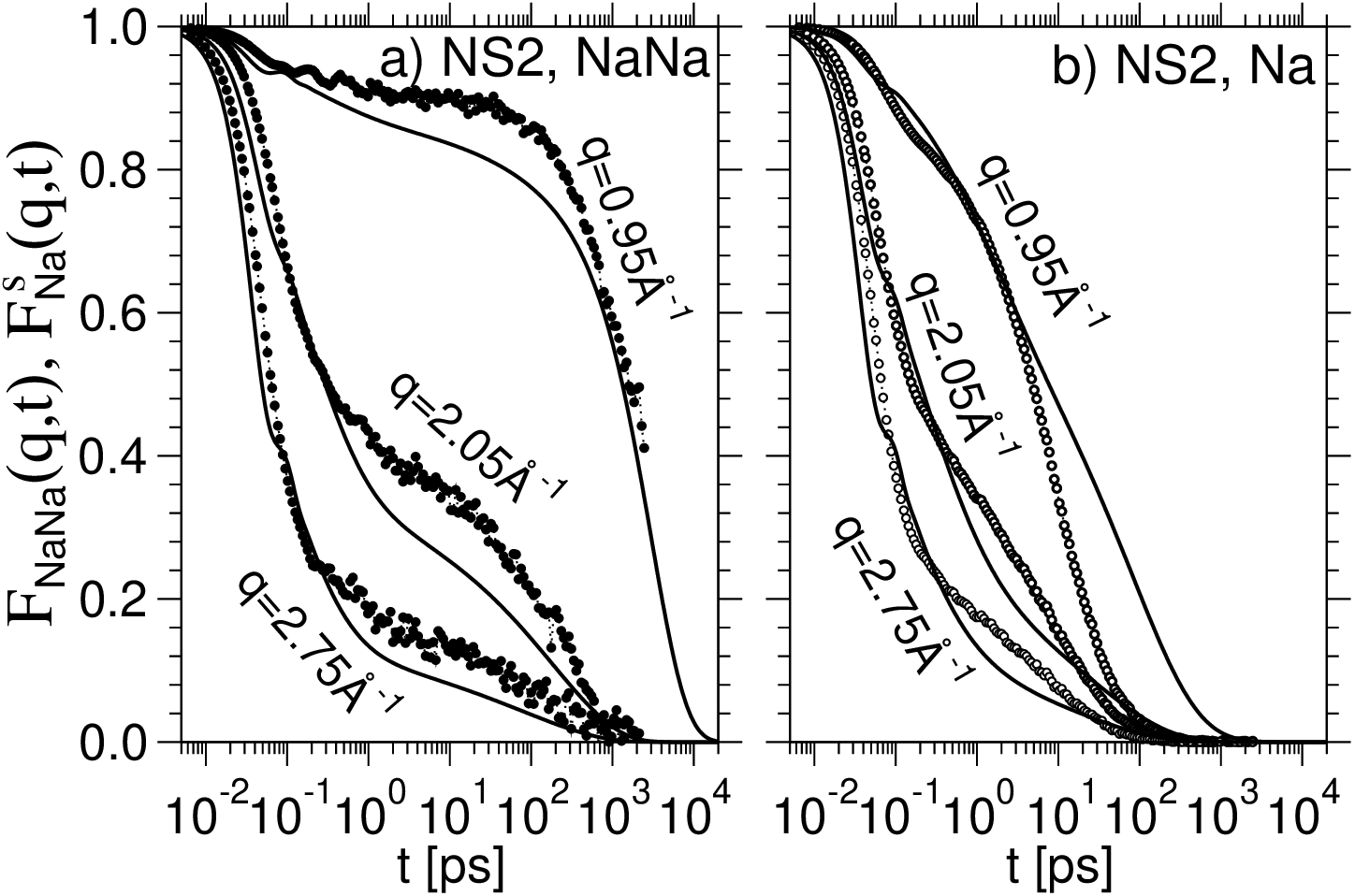}
\caption{\label{dyncomp}
Normalized sodium-sodium intermediate scattering functions,
$F_{\text{Na}\text{Na}}(q)$ (left panel), and sodium-self intermediate
scattering functions, $F^s_{\text{Na}}(q,t)$ (right panel), for sodium
disilicate (NS2). The symbols are MD results for $T=2100$\,K (from
Ref.~\protect\cite{horbach02}); the solid lines are MCT results for
$T_{\text{MCT}}=3410$\,K; $q$ values are indicated.
}
\end{figure}
The large difference in time scales between the matrix relaxation
and that of sodium transport, on almost all length scales,
is best evidenced in the sodium-sodium (coherent) and the
sodium-self (incoherent) correlation functions \cite{horbach02},
$F_{\alpha\alpha}(q,t)=S_{\alpha\alpha}(q,t)/S_{\alpha\alpha}(q)$ and
$F^s_\alpha(q,t)$ in the case $\alpha=\text{Na}$. The former follows
the slow matrix relaxation, while the latter decays much faster, giving rise
to fast sodium diffusion \cite{horbach02}. MD results at $T=2100\,\text{K}$
together with MCT results at $T_{\text{MCT}}=3410\,\text{K}$ for the two
quantities at selected wave numbers $q$ are shown in Fig.~\ref{dyncomp}.
We have chosen the latter temperatures for the comparison, since at
these temperatures the coherent relaxation times as obtained from MD
and MCT are very similar at $q=0.95\,\text{\AA}^{-1}$ (see also below).
At this wave number, the time-scale separation is most pronounced: while
the sodium-sodium correlations decay at $\tau\approx10^4\,\text{ps}$,
the sodium-self correlations decay at $\tau^s\approx10^2\,\text{ps}$,
i.e.\ two orders of magnitude faster. MCT underestimates this separation,
but it still gives a factor of about $50$ between the two time scales.
Note that in the hard-sphere system, the coherent and incoherent relaxation
times agree within a factor of $4$ at comparable intermediate $q$. Hence,
the theory captures the peculiarity of the fast sodium dynamics in NS2
qualitatively correctly. At the higher wave number $q=2.05\,\text{\AA}^{-1}$,
the time-scale separation is smaller, but still $\tau/\tau^s\approx20$ for the
MD data. The MCT result, $\tau/\tau^s\approx14$, here is even closer to the
simulation value.

Overall, the theory reproduces well the qualitative facts seen in the MD
simulations.  There are two major types of discrepancy visible: first, the
heights of the plateaus seen in the correlation functions at intermediate
times differ. They are connected to the glass form factors, for which
in the case of silica \cite{sciortino01} the agreement was quantitative
with static triplet correlations taken into account, but qualitative only
without them (as in our case). We therefore expect this first problem
to be curable by providing better static structure input. The second
problem is the shape of the correlators, especially the sodium-self
correlators at long times: for example, for $q=0.95\,\text{\AA}^{-1}$,
MCT predicts a long, extremely stretched decay from $0.6$ to zero over
more than 2 decades in time, which is not observed in the simulation.

\begin{figure}
\includegraphics[width=\figurewidth]{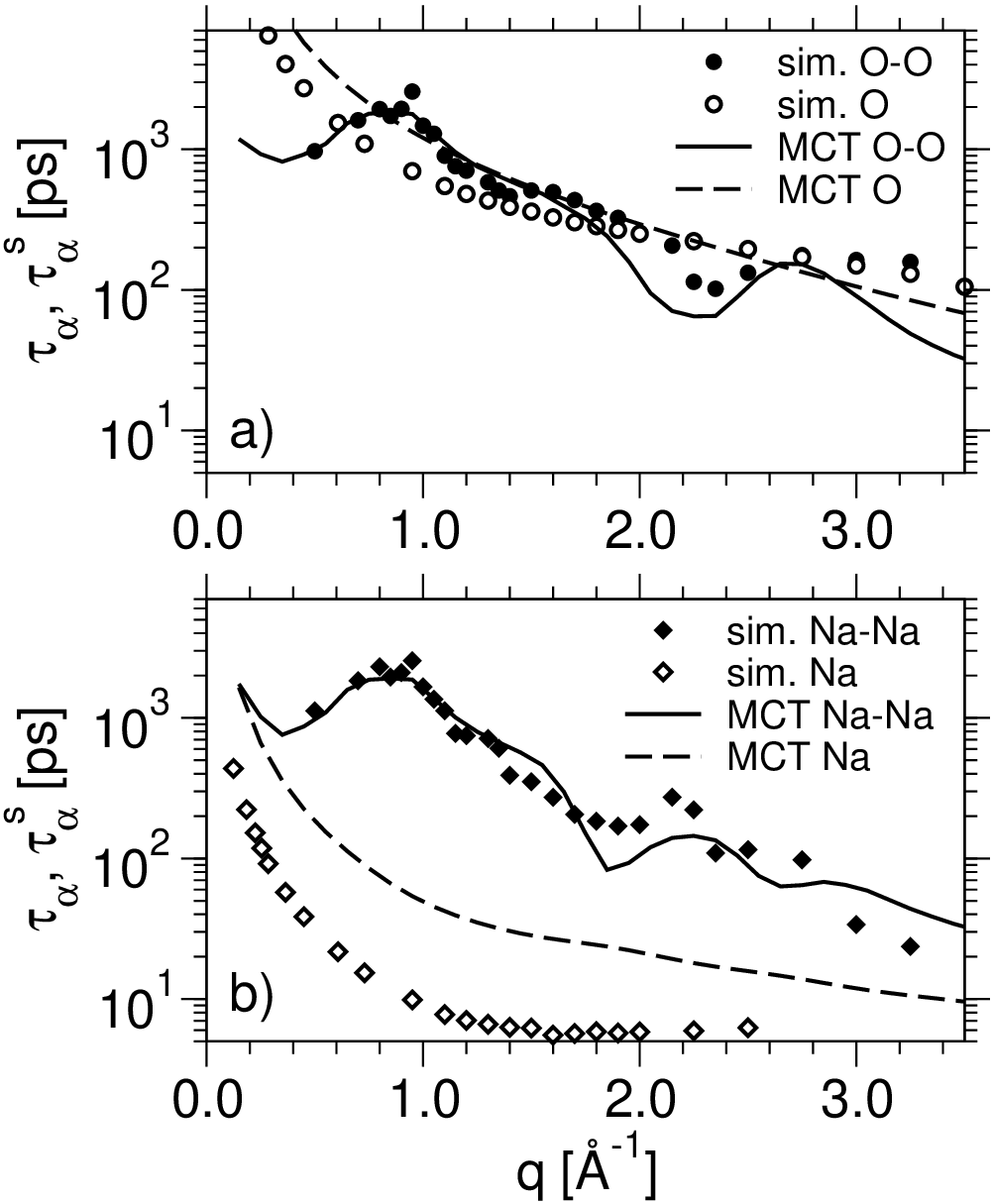}
\caption{\label{taualpha}
Structural relaxation times $\tau_\alpha(q)$ for coherent density
relaxation functions $F_{\alpha\alpha}(q,t)$ and $\tau^s_{\text{Na}}(q)$ 
for the incoherent density relaxation function $F^s_{\text{Na}}(q,t)$,
a) $\tau_{\text{O}}(q)$ and $\tau_{\text{O}}^s(q)$, 
b) $\tau_{\text{Na}}(q)$ and $\tau_{\text{Na}}^s(q)$.
As indicated, symbols are simulation results ($T=2100\,\text{K}$) 
and lines are MCT results ($T=3410\,\text{K}$). See text for the
definition of $\tau_\alpha(q)$ and $\tau^s_\alpha(q)$.
}
\end{figure}
To test the wave-vector dependence of the dynamics in more detail, we
turn to a discussion of structural relaxation times for the coherent,
$\tau_\alpha(q)$, and the incoherent correlators, $\tau^s_\alpha(q)$.
We define them through fits of stretched exponentials to the
correlation functions, $F_{\alpha\alpha}(q,t)\approx A_\alpha(q)
\exp[-(t/\tau_\alpha(q))^{\beta_\alpha(q)}]$ at long times,
with $\beta_\alpha(q)<1$ and $A_\alpha(q)<1$, and similarly for
$F_\alpha^s(q,t)$. This yields a reasonable description of the dynamics
for $t\gtrsim1\,\text{ps}$. In Fig.~\ref{taualpha}, coherent and
incoherent relaxation times are plotted for oxygen and sodium. MD and
MCT results are shown as symbols and lines, respectively. The coherent
relaxation times are, at fixed $q$, all of the same order of magnitude,
and they almost coincide around $q\approx 1\,\text{\AA}^{-1}$, the
region corresponding to the main structural features of the system
(note that this holds also for $\tau_{\text{Si}}(q)$). The values
obtained from MCT and those from MD are in good agreement given
that we only have one adjustable parameter. In particular, their
subtle $q$-dependence is reproduced. The oxygen-self
relaxation time $\tau_{\text{O}}^s(q)$ is of the same order as the
corresponding coherent one, $\tau_{\text{O}}(q)$, except for $q\to0$.
This is what one expects in a typical glass-forming liquid. On the
contrary, the sodium-self relaxation time
$\tau_{\text{Na}}^s(q)$ is much smaller than all other structural
relaxation times, both in the MCT and the MD results. Even if the MCT
results for this quantity are systematically higher than the MD data
(reflecting the difference in shape of the sodium self-correlators in
Fig.~\ref{dyncomp}), this confirms
that MCT indeed predicts the fast-ion dynamics in NS2 from the equilibrium
structure alone.

\begin{figure}
\includegraphics[width=\figurewidth]{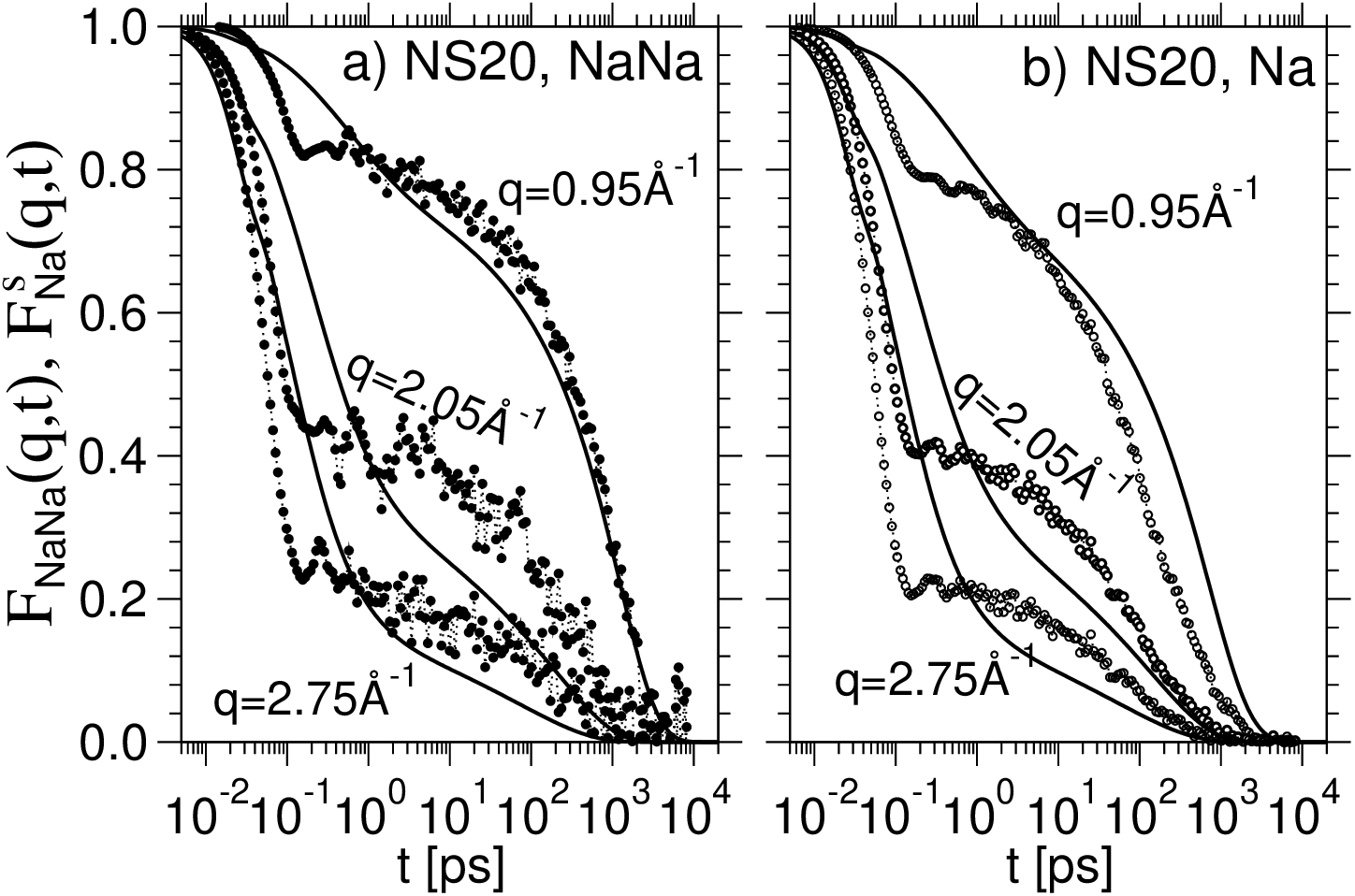}
\caption{\label{ns2ns20}
  Normalized sodium-sodium intermediate scattering functions,
  $F_{\text{Na}\text{Na}}(q)$ (left panel), and sodium-self intermediate
  scattering functions, $F^s_{\text{Na}}(q,t)$ (right panel), as in
  Fig.~\protect\ref{dyncomp}, but for NS20. MD results (symbols, from
  Ref.~\protect\cite{horbach03b}) are for $T=2500\,\text{K}$, MCT results
  (solid lines) for $T_{\text{MCT}}=4000\,\text{K}$.
}
\end{figure}
We now investigate the effect of sodium concentration on the dynamics by
discussing simulation and MCT results for NS20. Fig.~\ref{ns2ns20} shows
the coherent and incoherent sodium-density correlators for this system,
similar to the NS2 results shown in Fig.~\ref{dyncomp}.  For NS20, results
at higher $T$ than in NS2 are shown, in order to compare states which
show the same degree of slowing down, i.e.\ $\tau_{\text{Na}} ={\mathcal
O}(10^3\,\text{ps})$ for $q\approx0.95\,\text{\AA}^{-1}$.  The main effect
of reducing the sodium concentration from NS2 to NS20 on the dynamics is
that the sodium-self relaxation now is appreciably slower, as is evident
from the right-hand panel in Fig.~\ref{ns2ns20}.  The relaxation time
obtained from the MD data is $\tau_{\text{Na}}^s(q=0.95\,\text{\AA}^{-1})
\approx10^3\,\text{ps}$, similar to the $\tau_{\text{Na}}(q)$ value at
that $q$, i.e.\ all relaxation times, including that of the sodium-self
correlations, are now of the same order of magnitude at fixed $q$. MCT
again somewhat overestimates the sodium-self relaxation times, but the
qualitative sodium-concentration effect is well explained by the theory.

Also for NS20, the shape of the relaxation curves predicted by MCT for
the sodium-self correlators disagrees with that observed in MD. But here,
the discrepancy is found largely in the form of the relaxation towards
the plateau, while the further deviations visible in Fig.~\ref{ns2ns20}
are ascribable to the error in those plateau values as
explained above. The observation of deviations at relatively
short times, compared to the structural relaxation time, is typical
for comparisons between MCT and Newtonian dynamics MD simulations
\cite{nauroth97,foffi04}.

In summary, we have used MCT in conjunction with computer-simulated static
structure factors for different sodium silica melts.
The theory is able to explain the decoupling of time scales between
the sodium-self dynamics and the much slower network relaxation, as
well as the dependence of this decoupling on the sodium concentration.
The ratios of relaxation times are in qualitative agreement with the MD
results. Since MCT only makes use of the partial structure factors
$S_{\alpha\beta}(q)$, this shows that indeed a structure--dynamics relation
holds for the fast ion transport in these alkali silicates, and that MCT
provides a possible microscopic explanation for it.
The description is not yet on a quantitative level, but one can expect
that taking into account more detailed information about the equilibrium
static structure, i.e.\ static triplet correlations, will improve on at
least some of the deviations pointed out above.

\begin{acknowledgments}
We thank M.E.~Cates, W.~G\"otze, W.~Kob, and W.~Paul for their comments on the
manuscript. The authors
are supported through the Emmy Noether program of the DFG, grants
Ho~2231/2-1/2 (J.H.) and Vo~1270/1-1 (Th.V.), and EPSRC grant GR/S10377/01
(Th.V.).  Computing time on the JUMP at the NIC J\"ulich is gratefully
acknowledged.
\end{acknowledgments}


\bibliography{lit}
\bibliographystyle{apsrev}

\end{document}